\documentstyle[aps,prb,twocolumn,epsf,floats]{revtex}
\draft
\begin{document}
\wideabs{

\title{Nuclear spin relaxation 
       in integral and fractional quantum Hall systems}

\author{
   Izabela Szlufarska and Arkadiusz W\'ojs}
\address{
   Department of Physics,
   University of Tennessee, Knoxville, Tennessee 37996, \\
   and Institute of Physics,
   Wroclaw University of Technology, Wroclaw 50-370, Poland}

\author{
   John J. Quinn}
\address{
   Department of Physics,
   University of Tennessee, Knoxville, Tennessee 37996}

\maketitle

\begin{abstract}
We report on the numerical study of the relaxation rates of nuclear
spins coupled through the hyperfine interaction to a two dimensional
electron gas (2DEG) at magnetic fields corresponding to both fractional 
and integral Landau level (LL) fillings $\nu$.
The Hamiltonians of up to 20 interacting electrons are diagonalized
exactly in the spherical geometry, neglecting finite layer width,
disorder, and LL mixing.
The spectral functions $\tau^{-1}(E)$ describing response of the 2DEG
to the reversal of an embedded localized spin are calculated.
In a (locally) incompressible $\nu=1$ or ${1\over3}$ state, the finite
Coulomb energy of short spin waves, together with the small nuclear
Zeeman energy, prevent nuclear spin relaxation even in the limit of
vanishing electron Zeeman energy ($E_{\rm Z}$).
However, we find that the nuclear spins can couple to the internal
excitations of mobile finite-size skyrmions that appear in the 2DEG
at sufficiently low $E_{\rm Z}$ and at $\nu$ slightly different from
1 or ${1\over3}$.
The experimentally observed dependence of nuclear spin relaxation rate
on $E_{\rm Z}$ and $\nu$ is qualitatively explained in terms of the 
occurrence of skyrmions and antiskyrmions of various topological charge.
\end{abstract}
}

%%%%%%%%%%%%%%%%%%%%%%%%%%%%%%%%%%%%%%%%%%%%%%%%%%%%%%%%%%%%%%%%%%%%%
\section{Introduction}
%%%%%%%%%%%%%%%%%%%%%%%%%%%%%%%%%%%%%%%%%%%%%%%%%%%%%%%%%%%%%%%%%%%%%

Two decades ago, transport experiments on the two-dimensional 
electron gas (2DEG) systems in a high magnetic field $B$ revealed 
rich new physics associated with the unique properties of their 
charge excitations, including the pair of most striking phenomena, 
the integral\cite{Klitzing80,Laughlin81} and fractional
\cite{Tsui82,Laughlin83,Haldane83,Halperin84,Jain89} quantum 
Hall effects (IQHE and FQHE).
Both effects are the manifestation of a finite gap opening in the 
charge excitation spectrum at a series of (integral or fractional) 
values of the Landau level (LL) filling factor $\nu=1$, 2, ${1\over3}$,
${2\over3}$, etc., and of the quasiparticle nature of the elementary 
charge excitations of this series of gapped (and thus incompressible) 
ground states.\cite{Prange87,Chakraborty95}

Recent development of nuclear magnetic resonance\cite{Slichter90} 
(NMR) techniques allowed their successful application to the quantum 
Hall systems,\cite{Barret94,Tycko95,Kuzma98} and ultimately opened 
a new area of research associated with their spin degree of freedom.
The quantum Hall systems with spin excitations are very attractive 
to both theory and experiment because of their fundamental aspects 
as well as for potential for applications.
They are liquids with unique (Laughlin) correlations\cite{Laughlin83,%
Haldane87,parent} and unique excitations (integrally or fractionally 
charged quasiparticles without\cite{Laughlin83} or with\cite{%
Chakraborty86} spin, skyrmions\cite{Rajaraman82,Sondhi93,Fertig96,%
skyrmion} and their collective excitations,\cite{Brey95} 
and charge\cite{Haldane85} and spin waves\cite{Kallin84}).
On the other hand, the hyperfine coupling of the mobile electron 
spin excitations to the localized nuclear spins of the underlying 
atoms\cite{Berg90,Vagner88,Sinova00} creates the possibility of 
controlling the latter by inducing appropriate phase transitions in 
the 2DEG under variation of such experimentally adjusted macroscopic 
parameters as magnetic or electric fields, pressure, or temperature.
\cite{Hashimoto02}
If using nuclear spin states as a physical realization of quantum 
bits for storage of (and performing logical operations on) quantum 
information turned out to be possible, the coupled 2DEG-nuclei system 
should become a promising candidate for the spin memory elements 
of a quantum computer.\cite{Berman98}

The most enlightening experiments on spin quantum Hall systems were 
those that offered NMR evidence for the occurrence of skyrmions in 
the IQH\cite{Tycko95} and FQH\cite{Kuzma98} regime (confirmed by 
subsequent optical\cite{Davies98} and transport\cite{Hashimoto02,%
Kronmuller98} studies).
Skyrmions ($S_K^-$) and their conjugates, antiskyrmions ($S_K^+$), 
consist of $K$ neutral spin waves bound to a particle in the empty 
reversed-spin ($\uparrow$) LL or to a hole in the filled ($\downarrow$) 
LL, respectively.\cite{skyrmion}
In the IQH regime, the relevant particles are the reversed-spin 
electrons ($e_{\rm R}$) and the LL holes ($h$), and the skyrmions can 
be viewed as $S_K^-=e_{\rm R}(e_{\rm R}h)_K$ or $S_K^+=h(e_{\rm R}h)_K$ 
bound states, analogous to interband charged excitons $X_K^\pm$.
\cite{x-td}
In the FQH regime, skyrmions consist of reversed-spin quasielectrons
\cite{Chakraborty86} (QE$_{\rm R}$) and Laughlin quasiholes\cite{%
Laughlin83} (QH) bound to form $S_K^-={\rm QE}_{\rm R}({\rm QE}_{\rm R}
{\rm QH})_K$ and $S_K^+={\rm QH}({\rm QE}_{\rm R}{\rm QH})_K$.
\cite{skyrmion}
The analogy between FQH and IQH skyrmions is most evident in the 
composite fermion (CF) picture\cite{Jain89} in which QE$_{\rm R}$ 
and QH are represented by particles and holes at the integral filling 
of their effective CF LL's.
In both regimes, skyrmions are charged quasiparticles carrying large 
spin $K$. 
If the Zeeman energy $E_{\rm Z}$ in a sample is sufficiently small for
the isolated $e_{\rm R}$, $h$, QE$_{\rm R}$, and QH quasiparticles 
to become become unstable towards creation and binding of a number 
($K$) of spin waves to form skyrmions, the number of spin flips per 
particle added to or removed from a filled (electron or CF) LL is $K$.
This quantity ($K$) sets the slope of the electron spin polarization
$\left<S_z\right>$, proportional to the Knight shift measured in the 
NMR experiments,\cite{Tycko95,Kuzma98} as a function of the filling 
factor near $\nu=1$, ${1\over3}$, etc.

Of the most recent ones, particularly intriguing seems the NMR 
experiment of Kuzma {\sl et al.}\cite{Kuzma98} which revealed
extremely long nuclear spin relaxation time in the FQH regime, 
$\tau\le0.5$~s, exceeding times recorded earlier\cite{Davies98} 
by $\sim10^3$.
So different a relaxation time found in seemingly similar systems 
suggest that different microscopic mechanisms can be responsible 
for nuclear spin relaxation, depending on experimentally variable 
conditions.
Discussion of such mechanisms is the subject of this work.

We report on detailed numerical studies of the hyperfine interaction 
of the incompressible quantum Hall states at $\nu=1$ and ${1\over3}$, 
as well as their spin excitations (spin waves), reversed-spin 
quasiparticles, and skyrmions), with the localized nuclear spins.
The many-electron wave functions are obtained from 
exact-diagonalization calculations carried out in the Haldane 
spherical geometry, neglecting disorder and excitations to 
higher orbital LL's or to higher quantum well subbands.
The spectral function $\tau^{-1}(E)$ that describes response of 
the 2DEG to the reversal of an embedded localized spin and governs 
nuclear spin relaxation time $\tau$ for the particular microscopic 
2DEG-nucleus spin-flip process is calculated.

We find that in a incompressible $\nu=1$ or ${1\over3}$ state, 
the reversal of a nuclear spin creates a spin wave of a finite 
wave vector $k$ simply related to the area occupied by one electron, 
$(k\lambda)^2\approx\nu$ (where $\lambda$ is the magnetic length).
Since the spin wave dispersion $E_{\rm SW}(k)$ begins at the 
electronic Zeeman gap, $E_{\rm SW}(0)=E_{\rm Z}$, the energy of 
a spin wave coupled to a nuclear spin exceeds $E_{\rm Z}$ by a 
term $E_{\rm SW}(k)-E_{\rm SW}(0)$ which is of the order of the 
characteristic Coulomb energy $E_{\rm C}=e^2/\lambda\propto\sqrt{B}$.
Since $E_{\rm Z}$ and $E_{\rm C}$ are both much larger than
the nuclear Zeeman gap, the energy conservation is expected to 
exclude creation (or annihilation) of spin waves as an efficient 
mechanism for nuclear spin relaxation.
Consequently, very long relaxation times $\tau$ are expected 
for nuclear spins embedded in a (locally) incompressible IQH
or FQH state, even if $E_{\rm Z}$ could be made arbitrarily small
(by means of appropriate doping or application of pressure).

At $\nu$ slightly different from 1 or ${1\over3}$, skyrmions
(or reversed-spin quasiparticles) appear in the incompressible 
quantum Hall liquid.
The response function $\tau^{-1}(E)$ is calculated for these 
objects and shown to have peaks corresponding to their ``internal 
spin excitations'' in which the skyrmion spin $K$ and angular 
momentum $L$ both change by one ($S_K\rightarrow S_{K\pm1}$).
\cite{Fertig96,skyrmion} 
It is also found that the oscillator strength $\tau^{-1}_K$ 
of these transitions increases with increasing $K$.
Since the energy gap for the internal skyrmion excitations is 
much smaller than $E_{\rm Z}$ (and, in particular, it is equal
to the nuclear Zeeman gap at the series of values of $E_{\rm Z}$), 
the skyrmion--nucleus spin-flip processes will be allowed by the 
energy conservation law, and provide efficient nuclear spin 
relaxation mechanism under experimental conditions.

In both IQH and FQH regimes, our results imply critical dependence 
of the nuclear spin relaxation rate on the presence of skyrmions 
in the 2DEG (dependent on $\nu$, $E_{\rm Z}$, well width, etc.), 
in agreement with experiments.
However, it is quite remarkable that (because skyrmions of the $\nu
={1\over3}$ state occur only at much smaller values of $E_{\rm Z}$ 
than skyrmions at $\nu=1$) the energy required to create an electron 
spin excitation is much smaller in the IQH regime than in the FQH 
regime over a long range of $E_{\rm Z}$.
This is opposite to the relation between charge excitation gaps,
which scale as $\hbar\omega_c\propto B$ for IQH states and a (much
smaller) $E_{\rm C}\propto\sqrt{B}$ for FQH states.

At this stage of our study, we have limited ourselves to the 
calculation and analysis of a simple spectral function for the 
idealized many-electron states.
This only allows for estimates of {\em relative} relaxation times 
$\tau^{-1}$ for different microscopic processes, but not for their
actual magnitudes.
The model also neglects the effects of 
(i) the finite width of the quasi-2DEG, the tilt of the magnetic 
field, the particular density profile $\varrho(z)$ in the direction 
normal to the 2DEG plane, or the nuclear polarization profile 
$\varrho_{\rm N}(z)$ in that direction;\cite{Sinova00}
(ii) the interaction-induced electron or hole scattering to 
higher LL's (i.e., LL mixing);
(iii) disorder; or
(iv) non-equilibrium processes.
All these effects may become important in realistic experimental 
systems, leading to the reduction of the electronic interaction 
energy scale compared to the nuclear Zeeman energy due to the finite 
width of electronic wave functions (i) or screening (ii), dependence 
of the nuclear spin relaxation rate on the correlation between 
$\varrho(z)$ and $\varrho_{\rm N}(z)$ profiles (i), or localization 
of electronic excitations that relaxes the angular momentum 
conservation law in the spin-wave--nucleus or skyrmion--nucleus 
scattering processes (iii).
We plan to study these and other possible effects in the future, and 
the motivation for the analysis of the ideal model used here is based
on the fact that such model allows identification and classification 
of possible elementary microscopic spin-flip processes, as well as 
the formulation of the involved selection rules which will be only 
modified to a various degree depending on the specific experimental 
conditions.

%%%%%%%%%%%%%%%%%%%%%%%%%%%%%%%%%%%%%%%%%%%%%%%%%%%%%%%%%%%%%%%%%%%%%
\section{Model}
%%%%%%%%%%%%%%%%%%%%%%%%%%%%%%%%%%%%%%%%%%%%%%%%%%%%%%%%%%%%%%%%%%%%%

\subsection{Electron liquid}

As an extension of the earlier work on the spin excitations of the 
2DEG in the quantum Hall regime,\cite{Sondhi93,skyrmion} we study 
coupling of these excitations to the localized (e.g. nuclear) spins.
The model to describe the 2DEG is that of Ref.~\onlinecite{skyrmion},
except that it is now extended to include the presence of a nucleus.
In order to preserve the 2D symmetry of an infinite quantum well in 
a finite size calculation, the electrons are confined to a Haldane 
sphere\cite{Haldane83} of radius $R$.
The magnetic field $B$ normal to the surface is due to a Dirac
monopole in the center of the sphere.
The monopole strength $2Q$ is defined in the units of flux quantum
$\phi_0=hc/e$, so that $4\pi R^2B=2Q\phi_0$ and $\lambda=R/\sqrt{Q}$
is the magnetic length.
The single-electron states (monopole harmonics) are the eigenstates
\cite{Haldane83,Wu76} of magnitude ($l$) and projection ($m$) 
of angular momentum and of spin projection $\sigma$,
and they form $g$-fold ($g=2l+1$) degenerate LL's labeled by $n=l-Q$.

The cyclotron energy $\hbar\omega_c\propto B$ is assumed much
larger than the Coulomb energy $E_{\rm C}=e^2/\lambda\propto
\sqrt{B}$.
However, no assumption is made about the electron Zeeman energy,
and $\eta=E_{\rm Z}/E_{\rm C}$ is a (small) free parameter of
the model.
As a result, only the $\sigma=-{1\over2}$ ($\downarrow$) and 
$+{1\over2}$ ($\uparrow$) states of the lowest ($n=0$) LL need be 
included in the calculation, denoted simply by $\left|m\sigma\right>$.

The many-electron Hamiltonian in the lowest LL is
\begin{eqnarray}
\label{eqH}
   H &=& \sum
   c_{m_1\sigma}^\dagger
   c_{m_2\sigma'}^\dagger
	c_{m_3\sigma'}
   c_{m_4\sigma}
	\left<m_1m_2\left|V\right|m_3m_4\right>
\nonumber\\
   &+& \sum
   c_{m\uparrow}^\dagger
   c_{m\uparrow}
   E_{\rm Z},
\end{eqnarray}
where $c_{m\sigma}^\dagger$ and $c_{m\sigma}$ are the electron
creation and annihilation operators, the summation goes
over all orbital and spin indices, and $V$ is the Coulomb
interaction potential.
The $N$-electron eigenstates are expanded in the basis of Slater
determinants
\begin{equation}
\label{eqbas}
   \left|m_1\sigma_1\ldots m_N\sigma_N\right> =
   c_{m_1\sigma_1}^\dagger	\ldots
   c_{m_N\sigma_N}^\dagger
   \left|{\rm vac}\right>,
\end{equation}
where $\left|{\rm vac}\right>$ is the vacuum state.
Basis (\ref{eqbas}) allows automatic resolution of two good
many-body quantum numbers, projections of spin ($S_z=\sum\sigma_i$)
and of angular momentum ($L_z=\sum m_i$).
However, the lengths of spin ($S$) and of angular momentum ($L$)
are resolved numerically in the numerical diagonalization of each
$(S_z,L_z)$ Hilbert subspace.
The additional quantum number $K={1\over2}N-S$ measures the number 
of reversed spins relative to the maximally polarized state.
The many-electron states on a (finite) sphere converge to the
states on an (infinite) plane in the $Q=(R/\lambda)^2\rightarrow
\infty$ limit, only the spherical orbital numbers $L$ and $L_z$
must be appropriately\cite{Avron78} replaced by the planar ones, 
the projections of total and center-of-mass angular momentum,
$M$ and $M_{\rm CM}$.

In the lowest energy states of the system described by Hamiltonian
(\ref{eqH}) near the integral or odd-denominator fractional filling 
of the lowest LL ($\nu=1$ or ${1\over3}$), a small number of spin 
waves or skyrmions move (to a good approximation, independently) 
in the appropriate incompressible quantum Hall ``background'' state.
Being charge neutral excitations, spin waves move along straight 
lines and carry linear momentum $\hbar k$.
On a sphere, their linear orbits are closed into great circles,
and the linear wave vector $k$ takes on discrete values following 
from the quantization of angular momentum, $L=kR=0$, 1, 2, \dots.
Skyrmions ($S_K$) on the other hand are charged, particle-like 
excitations that move along circular cyclotron orbits similar to 
those of electrons.
Their motion is therefore similarly quantized in both geometries, 
with the lowest LL of states having $M=M_{\rm CM}+K$ and $M_{\rm CM}
=0$, $\pm1$, $\pm2$, \dots\ (on a plane) or $L=Q-K$ and $L_z=L$, 
$L-1$, $L-2$, \dots\ (on a sphere), respectively.

Both spin waves and skyrmions may become localized in the presence
of disorder which has been ignored in this work.
However, the dominant effects of such localization, at least in the 
weak disorder regime, can easily be predicted by analogy with the 
interband emission of neutral and charged excitons.
For spin waves, the localization alters the relative occupation of
different $k$-states (in particular, it increases occupation of the 
$k=0$ state).
However, we show later that spin waves do not couple efficiently to 
nuclear spins regardless of $k$.
For skyrmions, localization of their cyclotron orbits in the lower 
energy areas (without significant distortion of their wave functions)
has two consequences:
(i) Freezing of the positions of (few) skyrmions relative to (many) 
nuclei and thus variation of the electron spin polarization from one 
nucleus to another; this causes broadening of the Knight shift over 
the measured sample in the NMR experiment; delocalization of skyrmions 
and restoration of uniform electron spin polarization at higher 
temperatures is called the ``motional narrowing'' of NMR lines.
(ii) Variation of skyrmion energies due to confinement relaxes the 
energy conservation law for the skyrmion--nucleus spin-flip processes,
and thus broadens the minima of the nuclear spin relaxation time 
$\tau$ as a function of $E_{\rm Z}$.
Nevertheless, our main conclusions regarding the form of the 
continuous (due to spin waves) and discrete (due to skyrmions)
parts of the 2DEG response function, and the role of the two types
of spin excitations for nuclear spin relaxation remain valid 
independently of localization.

\subsection{Coupling to nuclear spin}

The coupling of the electron system (the ``background'' quantum Hall 
state and its spin excitations) to a single isolated localized 
nuclear spin will be described by the contact hyperfine interaction
Hamiltonian\cite{Slichter90}
\begin{equation}
   F = A \sum_{j,k}\,{\bf I}_j {\bf S}_k\,
	\delta\left({\bf r}_j-{\bf R}_k\right),
\end{equation}
where $A$ is the coupling constant, and ${\bf S}_j$ and ${\bf I}_k$ 
(${\bf r}_j$ and ${\bf R}_k$) denote the spin (position) of the $j$th 
electron and $k$th nucleus, respectively.
Moreover, the distance between nearest nuclei will be assumed 
sufficiently large to justify neglecting their direct dipolar 
interaction, and exclude any multi-nucleus phenomena.

Due to the translational/rotational invariance of the 2DEG, the 
position of the nucleus can be conveniently chosen at the north 
pole of the sphere, where all electron wave functions of the 
lowest LL vanish, except for $\left|l\uparrow\right>$ and $\left|
l\downarrow\right>$.
Then, ignoring the overall coupling constant (independent of the 
system size, $R$ or $Q$), the transverse part of $F$ describing
the spin-flip processes with $\Delta S_z=1$ and projected onto 
the lowest LL simplifies to
\begin{equation}
\label{eqF}
   F = c_{l\uparrow}^\dagger c_{l\downarrow}.
\end{equation}
Clearly, the reversal of the localized (nuclear) spin is accompanied 
by a reversal of an electronic spin spread over a (cyclotron) orbit 
of finite radius $\sim\lambda$.

According to the Fermi golden rule, the oscillator strength 
$\tau^{-1}_{if}$ for the transition between a given pair of 
initial and final electronic eigenstates, $\left|i\right>$ 
and $\left|f\right>$, is proportional to the square of the 
matrix element of $F$,
\begin{equation}
   \tau^{-1}_{if}=\left|\left<f\right|F\left|i\right>\right|^2,
\end{equation}
and, accordingly, the spectral function for a given initial (ground) 
state is
\begin{equation}
\label{eqtau}
   \tau^{-1}(E)\equiv\tau^{-1}_i(E)
      =\sum_f \tau^{-1}_{if}\delta[E-(E_i-E_f)], 
\end{equation}
where $E_i$ and $E_f$ are the energies of the initial and final
states, respectively.
Also, from Eq.~(\ref{eqF}), $F$ couples the electron states with 
equal $L_z$ and $S_z$ different by one.

Note that $\tau^{-1}(E)$ is only defined up to a coupling constant
that we are unable to calculate.
Therefore, when comparing our results with experiment, only the 
relative intensities of different microscopic spin-flip processes
(e.g., processes that involve spin waves with different $k$, 
skyrmions with different $K$, or skyrmions as opposed to spin waves) 
are meaningful, but we are unable to estimate the actual magnitude 
of the nuclear spin relaxation rates.
However, only these relative intensities are a universal property
of the electron quantum Hall system, virtually independent of many
experimentally variable parameters (electron density, nuclear spin 
polarization profile across the 2DEG plane\cite{Sinova00}, etc.).
While spectral function (\ref{eqtau}) captures the essential physics
of the interaction between an (ideal) Laughlin liquid and localized
spins, the details of both the liquid and the spins must be included
in a realistic calculation of the relaxation rates.

Due to the simple form of the operator $F$ in our basis (\ref{eqbas}),
the oscillator strengths can be easily evaluated for any known pair 
of $\left|i\right>$ and $\left|f\right>$ eigenstates.
Therefore, the most difficult computational task is the accurate 
calculation of the many-electron eigenfunctions.
Moreover, because of the breaking of both spatial and spin symmetry
by the operator $F$, entire multiplets with different $L_z$ and $S_z$ 
must be calculated for each $L$ and $S$.
Hence, the proper identification of the relevant many-electron 
eigenstates that:
(i) describe a planar system in the $R/\lambda\rightarrow\infty$ limit,
(ii) have significant $\tau^{-1}$, 
becomes essential.

%%%%%%%%%%%%%%%%%%%%%%%%%%%%%%%%%%%%%%%%%%%%%%%%%%%%%%%%%%%%%%%%%%%%%
\section{Integral quantum Hall regime}
%%%%%%%%%%%%%%%%%%%%%%%%%%%%%%%%%%%%%%%%%%%%%%%%%%%%%%%%%%%%%%%%%%%%%

\subsection{Spin waves}

We begin with the integral quantum Hall regime and the filling 
factor of precisely $\nu=1$.
We evaluate numerically the oscillator strengths $\tau^{-1}$ 
for all possible transitions induced by the operator $F$ from 
the initial nondegenerate incompressible IQH ground state with 
$L=0$, $S_z=-{1\over2}N$, and $K=0$.
From the commutator $[F,S^2]$ it can easily be shown that the 
spin-flip transition defined by $F$ couples the $K=0$ initial
state $\left|i\right>$ to two different subspaces, with $K=0$ or 1.
However, the length of the total projection of $F\left|i\right>$,
the vector obtained by acting by $F$ on the initial IQH ground state, 
onto the $K=0$ subspace equals $N^{-1}$ in a finite $N$-electron 
initial state, and thus it disappears in the $N\rightarrow\infty$ 
limit.
Therefore, not surprisingly, the only spin excitations coupled to 
an infinite (planar) $\nu=1$ state by $F$ are those with $K=1$.
Of these, the only ones with significant oscillator strength
$\tau^{-1}$ turn out to be the spin wave states, which at the 
same time are the lowest energy excitations at $\nu=1$.

\begin{figure}[t]
\epsfxsize=3.40in
\epsffile{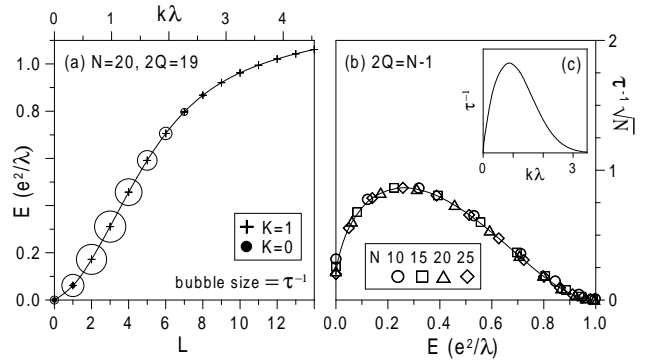}
\caption{
   (a) 
   The spin wave energy spectrum (energy $E$ vs.\ angular momentum 
   $L$ and wave vector $k$) at $\nu=1$ calculated for $N=20$ 
   electrons on Haldane sphere.
   The bubble diameters give the oscillator strength $\tau^{-1}$ for 
   the spin wave emission coupled to the nuclear spin reversal. 
   (b) 
   The response function of a planar $\nu=1$ state to a nuclear spin 
   reversal (oscillator strength $\tau^{-1}$ calculated on a sphere 
   for different $N\le25$ and normalized by $\sqrt{N}$ vs.\ energy $E$).
   (c)
   The response function $\tau^{-1}$ shown as a function of wave 
   vector $k$.
   Zeeman energy $E_{\rm Z}$ is excluded and $\lambda$ is the 
   magnetic length.}
\label{fig1}
\end{figure}

The numerical results for the spin-flip transitions corresponding 
to creation (annihilation) of spin waves in a finite $\nu=1$ state 
of $N=20$ electrons at $2Q=N-1=19$ are shown in Fig.~\ref{fig1}(a).
In this and all other spectra in the paper $E$ stands for the energy 
difference between the final and initial states and is given in the 
units of $E_{\rm C}=e^2/\lambda$.
The Zeeman energy $E_{\rm Z}$ is not included.
The horizontal axis shows the total angular momentum $L$ (for the
spin waves, $L=kR$), and the oscillator strength $\tau^{-1}$ is 
proportional to the diameter of a bubble around each energy level
marked with a cross.

As could be predicted from Eq.~(\ref{eqF}), the reversal of an 
electronic spin induced by $F$ occurs over an area corresponding 
to a cyclotron orbit.
This sets a characteristic length scale $\xi\sim\lambda$ for the 
efficient spin-flip process, and indeed in Fig.~\ref{fig1}(a) 
$\tau^{-1}$ has a maximum at a finite $L$, while it vanishes 
in the limits of both small and large $L$.

To determine the spectral function of a $\nu=1$ state of an infinite 
2DEG we have compared data obtained for different electron numbers, 
$N\leq25$, and plotted the results together in Fig.~\ref{fig1}(b).
The oscillator strengths for discrete values of $E$ are multiplied 
by the factor $\sqrt{N}\propto\sqrt{Q}\propto R/\lambda$, which comes 
from normalization of the wave function of the extended spin wave 
over the entire sphere.
All data points lie nicely on one regular curve that describes the 
spin wave creation/annihilation in both finite (spherical) and infinite 
(planar) systems.
As it is the characteristic wave vector $k$ (through the characteristic 
length scale $\xi$) rather than the energy $E$ that determines the
position of the maximum of $\tau^{-1}$, in the inset (c) we replot it
as a function of $k$ (only setting $\tau^{-1}(0)$ to zero as appropriate
for an infinite system).
$\tau^{-1}(k)$ is a more universal characteristic of the
$\nu=1$ IQH state than $\tau^{-1}(E)$ in a sense that the spin wave 
dispersion $E_{\rm SW}(k)$, in an ideal 2D system derived by Kallin 
and Halperin,\cite{Kallin84} in experiment may also depend on 
additional characteristics of actual 2DEG (e.g., the well width).
As expected, $\tau^{-1}(k)$ has a maximum at $k\sim\lambda^{-1}$,
which defines the spin-flip length scale of $\xi\sim\lambda\propto
\sqrt{B}$.

Based on Fig.~\ref{fig1}, we make the following observations:
(i) The incompressible $\nu=1$ liquid responds to the 
reversal of a localized spin by emission of a spin wave whose 
kinetic energy, $E_{\rm SW}(k)$, increases as a function of $k$;
(ii) The response function $\tau^{-1}(E)$ vanishes in both $k=0$ 
and $\infty$ limits, and it reaches maximum at $E$ corresponding 
to $k\sim\lambda^{-1}$;
(iii) When electron Zeeman energy is added, the energy of a 
$k\sim\lambda^{-1}$ spin wave that can couple to a localized spin 
reversal is a sum of two terms, $E_{\rm Z}$ and $E_{\rm SW}
(\lambda^{-1})$; and
(iv) Since $E_{\rm SW}(\lambda^{-1})\approx{1\over4}E_{\rm C}$ 
(in an ideal system\cite{Kallin84}) is typically much larger 
than the nuclear Zeeman energy, the energy conservation prevents 
efficient relaxation through emission of spin waves in a (locally) 
incompressible liquid regardless of the value of $E_{\rm Z}$.

This weak coupling of the $\nu=1$ state to the nuclear spins obtained
above agrees well with long nuclear spin relaxation times measured at 
this filling factor at low temperatures.\cite{Tycko95}
It is evident from the experiments showing rapid increase of the 
relaxation rate when either $\nu$ is moved away from 1 or temperature 
is elevated (from 2.1 to 4.2~K)\cite{Tycko95} that charged excitations 
provide more efficient mechanism for nuclear spin relaxation than the 
spin waves.
Let us then move on to the analysis of the spin-flip processes in 
the presence of such excitations, reversed-spin quasiparticles and 
skyrmions.

\subsection{Skyrmions}

\begin{figure}[t]
\epsfxsize=3.40in
\epsffile{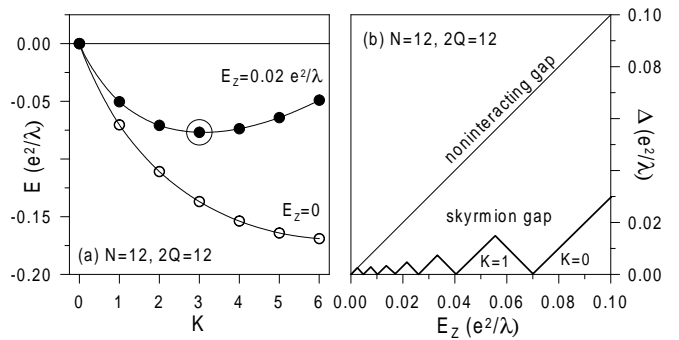}
\caption{
   (a) 
   The skyrmion energy spectrum (energy $E$ vs.\ reversed spin number 
   $K$) at $\nu=1^\pm$ calculated for $N=12$ electrons on Haldane sphere.
   Open and full symbols correspond to the Zeeman energy $E_{\rm Z}=0$
   and $0.02\,e^2/\lambda$, respectively ($\lambda$ is the magnetic 
   length).
   (b) 
   The energy gap $\Delta$ for skyrmion spin excitations 
   ($S_K\rightarrow S_{K\pm1}$) at $\nu=1^\pm$ as a function of 
   $E_{\rm Z}$, compared to the spin wave gap $\Delta=E_{\rm Z}$
   at $\nu=1$.}
\label{fig2}
\end{figure}

It is well-known that an extra particle (a reversed-spin electron or a 
spin hole) added to the $\nu=1$ ferromagnetic ground state induces and 
binds spin waves (whose number $K$ depends on $E_{\rm Z}/E_{\rm C}$)
to form a skyrmion\cite{Sondhi93} ($S_K$), a particle-like charged 
excitations carrying (possibly large) spin $K$.
In Fig.~\ref{fig2}(a) we replot the (anti)skyrmion energy spectrum 
calculated earlier in Ref.~\onlinecite{skyrmion} for $N=12$ electrons 
at $2Q=12$.
Using the exact particle-hole symmetry within an isolated LL, this 
state can be mapped onto one in which an extra reversed-spin electron 
is added to the $\nu=1$ state at the same value of $2Q=12$, and hence 
it will be denoted here as $\nu=1^\pm$.
Again, the vertical axis gives the skyrmion energy $E$ measured from 
the maximally polarized ($K=0$) state which in this case corresponds 
to one spin hole in the $\nu=1$ state.
On the horizontal axis we show the skyrmion spin (or size), $K$, 
which is also related\cite{skyrmion} to its angular momentum, $L=Q-K$.

The open symbols in Fig.~\ref{fig2}(a) mark the skyrmion energy 
spectrum $E_{\rm S}(K)$, excluding the Zeeman energy.
In an infinite system (and not only at $\nu=1$, but also in the FQH 
regime discussed in the following section), it can be quite accurately 
approximated by $E_{\rm S}(K)\approx-{\cal E}\,[K/(K+1)]^\alpha$.
In an ideal 2D system, the binding energy of an infinite skyrmion 
at $\nu=1$ is known exactly,\cite{Sondhi93} ${\cal E}={1\over4}
\sqrt{\pi/2}\,E_{\rm C}$, and any choice of $\alpha\sim1$ captures 
the most essential feature of $E_{\rm S}(K)$, which is that $E_{\rm S}
(K-1)-E_{\rm S}(K)>E_{\rm S}(K)-E_{\rm S}(K+1)$ for each $K$.

Although our numerics yields $\alpha\approx1.7$ at $\nu=1$, it is 
quite illuminating to solve the simplest case of $\alpha=1$ (the 
equally simple arithmetics for $\alpha=2$ gives essentially identical 
answer; moreover, for $\nu={1\over3}$ it actually seems that $\alpha
\approx1$).
When the Zeeman term $KE_{\rm Z}$ is added to the skyrmion energy as 
shown with full dots in Fig.~\ref{fig2}(a) the ground state becomes 
a finite-size skyrmion with a certain $K$ (as marked with an open 
circle, for $E_{\rm Z}=0.02\,E_{\rm C}$ it turns out to be $K=3$).
Using our simple model
\begin{equation}
\label{eqES}
   E_{\rm S}(K)=-{\cal E}{K\over K+1}+KE_{\rm Z}
\end{equation}
we obtain that the transition between the $S_{K-1}$ and $S_K$ ground 
states occurs at $E_{\rm Z}={\cal E}/[K(K+1)]$ (in particular, the 
smallest skyrmion, $S_1$, is stable below $E_{\rm Z}={1\over2}{\cal 
E}$), and the excitation gap $\Delta_K$ from $S_K$ to the lower of 
the $S_{K-1}$ or $S_{K+1}$ states [it is plotted in Fig.~\ref{fig2}(b) 
for the numerical data of Fig.~\ref{fig2}(a)] reaches its maximum 
value of $\Delta_K={\cal E}/[K(K+1)(K+2)]$ at $E_{\rm Z}={\cal E}/
[K(K+2)]$.

As seen in Fig.~\ref{fig2}(b), the most striking feature of the 
skyrmion energy spectrum in the presence of the (sufficiently small) 
Zeeman energy is that, in contrast to the spin wave spectrum of the
(locally) incompressible IQH liquid, the gap for spin excitations is 
greatly reduced compared to (and largely independent of) $E_{\rm Z}$.
Indeed, it follows from our expression for the maximum of $\Delta_K$ 
that $\Delta_K/E_{\rm Z}\le(K+1)^{-1}$ and $\Delta_K/E_{\rm Z}\le
\sqrt{E_{\rm Z}/{\cal E}}$. 
Moreover, the gap skyrmion $\Delta_K$ can be brought to resonance 
with an arbitrarily small nuclear Zeeman energy at the at entire 
series of $E_{\rm Z}$ near the $S_{K-1}\leftrightarrow S_K$ 
transitions.

Being charged objects, skyrmions move along electron-like cyclotron 
orbits and repel one another through an effective short-range 
pseudopotential similar to that of electrons in the lowest LL.
\cite{skyrmion}
Such short-range repulsion causes Laughlin correlations between the 
skyrmions,\cite{parent} which therefore avoid high energy collisions 
with one another and behave as well-defined quasiparticles moving 
independently in the underlying IQH liquid.
We know that in a similar, electron--valence-hole system, Laughlin 
correlations of interband charged excitons ($X^-$) with one another 
as well as with surrounding electrons simplify the photoluminescence 
of a dilute ($\nu<{1\over3}$) 2DEG to the recombination of spatially 
isolated $X^-$'s.\cite{x-correl}
By analogy, we expect that the many-skyrmion effects can be excluded
from the skyrmion--nucleus spin-flip scattering.

To verify the possibility of efficient nuclear spin relaxation through 
the interaction with skyrmions, we have evaluated the spectral functions
$\tau_i(E)$ for the initial state $\left|i\right>$ corresponding to 
a single skyrmion (or antiskyrmion) in a finite, $N$-electron $\nu=1$ 
system.
Due to the finite size of the skyrmion, its ground state is a 
degenerate electron-like LL.
On Haldane sphere, this LL is represented by the angular momentum
multiplet at $L=Q-K$.
Different values of $L_z=-L$, $-L+1$, \dots, $L$ label different 
cyclotron orbits, and the closest orbit to the north pole (the
position of the nucleus) is that with $L_z=L$ for a skyrmion ($S_K^-$) 
and $L_z=-L$ for an antiskyrmion ($S_K^+$).
Clearly, the value of $\tau_{if}$ for a given $i\rightarrow f$ 
transition involving a skyrmion depends on $L_z$, which plays
the role of an impact parameter of the skyrmion--nucleus collision.

\begin{figure}[t]
\epsfxsize=3.40in
\epsffile{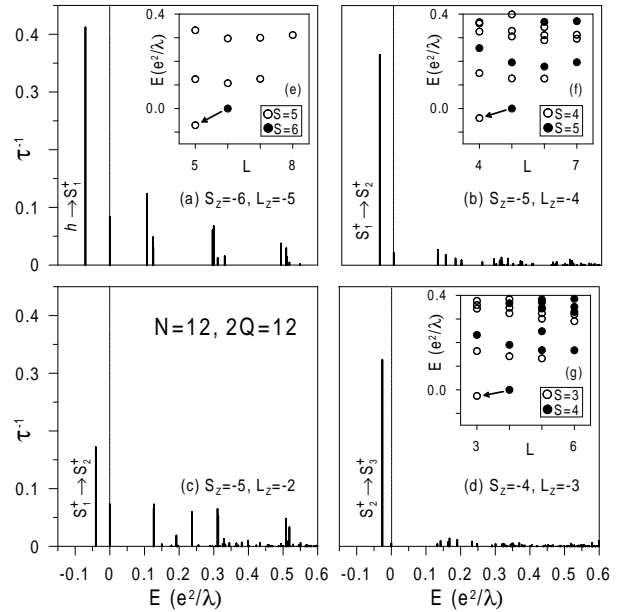}
\caption{
   The spectral function of the hyperfine transition operator $F$ 
   (oscillator strength $\tau^{-1}$ vs.\ energy $E$) at $\nu=1^\pm$ 
   calculated for $N=12$ electrons on Haldane sphere.
   Strong peaks at $E<0$ correspond to the ``internal'' skyrmion
   transitions $S_K\rightarrow S_{K+1}$ indicated with arrows in
   the energy spectra shown in the insets (e--g).
   $L_z$ is the skyrmion angular momentum projection related to the 
   impact parameter of the skyrmion--nucleus collision.
   Different frames correspond to different initial electron states:
   $h$ (a), $S_1^+$ (b), and $S_2^+$ (d) close to the nucleus, 
   and $S_1^+$ farther from the nucleus (c).
   Zeeman energy $E_{\rm Z}$ is excluded and $\lambda$ is the 
   magnetic length.}
\label{fig3}
\end{figure}

We have studied numerically a few systems with different values of 
$N$ and $2Q=N$. 
For each $N$ we have calculated the complete $\tau_{if}$ spectra 
corresponding to the (anti)skyrmionic initial states $\left|i\right>
=S_0^+$, $S_1^+$, $S_2^+$, \dots\ with different values of $L_z$,
and to all possible finite states $\left|f\right>$.
As an example, in Fig.~\ref{fig3} we show the results for $N=12$
and the initial states with $K=0$, 1, and 2 (for $K=1$ data for 
two values of $L_z$ is shown).
The transition energy given on the horizontal axis is $E=E_f-E_i$ 
(excluding $E_{\rm Z}$), and the units of $\tau^{-1}$ on the vertical
axis follow from Eq.~(\ref{eqF}).
In the insets we display the corresponding energy spectra in which 
the initial skyrmion states as well as all the final states can be 
identified for each transition.

All the spectra are quite similar in that they can be decomposed
into the quasi-continuous part at $E>0$ (here discrete because of 
size quantization) due to the response of the underlying IQH state
discussed in the preceding subsection, and a single peak at $E<0$
due to a $S_K\rightarrow S_{K+1}$ transition.
Since we are mostly interested in the processes that may conserve 
energy to allow efficient nuclear spin relaxation, let us neglect
all $E>0$ transitions and concentrate on the $S_K\rightarrow S_{K+1}$ 
one, whose energy including $E_{\rm Z}$ can be made equal to the 
nuclear Zeeman gap.
Clearly, it is only allowed for $|L_z|\le Q-(K+1)$ and its intensity 
quickly decreases when $L_z$ is increased from the minimum allowed 
value (such increase of $L_z$ corresponding to an increase of the 
nucleus--skyrmion average separation before the collision, i.e.\ 
of the impact parameter).
Because we assume no localization and thus allow that a skyrmion 
moves freely over the position of the nucleus (placed at the north 
pole by an arbitrary choice), it is more physical to consider 
$\tau^{-1}$ summed over all allowed values of $L_z$ as 
a characteristic of this ``internal skyrmion transition.''

\begin{figure}[t]
\epsfxsize=3.40in
\epsffile{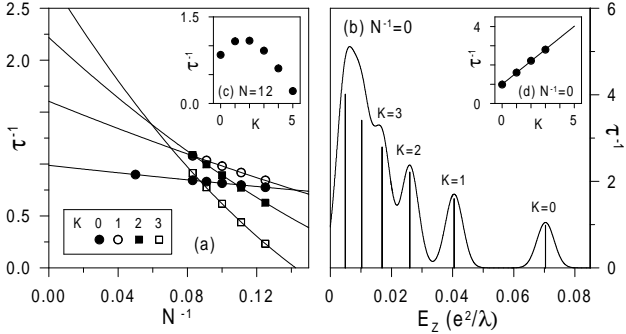}
\caption{
   (a) 
   The oscillator strengths $\tau^{-1}$ of the ``internal'' skyrmion 
   transitions $S_K\rightarrow S_{K+1}$ induced by a nuclear spin 
   reversal, as a function of the inverse electron number, $N^{-1}$.
   (b) 
   Same $\tau^{-1}$, but calculated in a finite, $N=12$ electron 
   system and plotted as a function of the skyrmion spin $K$.
   (c) 
   Same $\tau^{-1}$, but extrapolated to $N\rightarrow\infty$ and
   plotted as a function of the Zeeman energy $E_{\rm Z}$.
   The position of each peak is the value of $E_{\rm Z}$ at which
   the transition energy $\Delta$ is zero.
   The solid line includes Gaussian broadening.
   (d) Same as (b) but for data extrapolated to $N\rightarrow\infty$.}
\label{fig4}
\end{figure}

In order to complete the analysis of the role of skyrmions in 
nuclear spin relaxation, to the facts that skyrmion--nucleus spin 
scattering has finite oscillator strength and that it can conserve 
energy, we ought to add the dependence on the skyrmion size, $K$.
In Fig.~\ref{fig4}(b) we display the total (summed over all $L_z$) 
values of $\tau^{-1}$ for the $S_K\rightarrow S_{K+1}$ transitions
calculated for $N=12$ and plotted as a function of $K$.
Surprisingly, the function $\tau^{-1}(K)$ has a maximum at a finite $K$.
In order to estimate $\tau^{-1}$ in an infinite system, we have 
recalculated $\tau^{-1}(K)$ for different $N$.
The results are shown in Fig.~\ref{fig4}(a) where we plot $\tau^{-1}$
for $K=0$, 1, 2, and 3 as a function of the inverse system size, 
$N^{-1}$.
Very regular dependence of $\tau^{-1}$ on $N^{-1}$ for each $K$ 
allows their accurate (quadratic) extrapolation to the $N^{-1}
\rightarrow0$ limit, as indicated with the solid lines.
The result of the extrapolation is that in an infinite (planar) 
system, $\tau^{-1}$ increases monotonously with increasing $K$,
which indicates that nonmonotonic behavior in Fig.~\ref{fig4}(b) 
is an artifact.

The nearly linear increase of the extrapolated values of $\tau^{-1}$ 
with increasing $K$ shown in Fig.~\ref{fig4}(d) suggests that 
the total intensity $\tau^{-1}$ (summed over all $L_z$) depends 
predominantly on the skyrmion area.
Because in experiment the differential cross-section for the
skyrmion--nucleus collisions also depend on the skyrmion area,
one can expect that also the nuclear relaxation rate will increase 
as a function of $K$ (at a constant number of skyrmions).
In Fig.~\ref{fig4}(c) we plot the $\tau^{-1}$ peaks corresponding 
to subsequent $S_K\rightarrow S_{K+1}$ transitions as a function
of $E_{\rm Z}$ at which the energy of this transition ($E$) vanishes 
(compare Fig.~\ref{fig2}(b) in which the closing of the gap is shown).
Assuming that the skyrmion--nucleus spin scattering is a dominant
nuclear spin relaxation process and that it is most efficient when 
$E\approx0$, the curves obtained by broadening of the discrete peaks
with Gaussians imitate the nuclear spin relaxation rate as a function 
of $E_{\rm Z}$.
It is remarkable that when the value of $E_{\rm Z}$ is lowered,
the peaks with higher $K$ are selected from the spectral function,
the separation between the neighboring peaks decreases and their 
intensity increases.
Let us stress that this expected behavior for $\nu=1^\pm$ differs
qualitatively from what we predict at precisely $\nu=1$, where 
the relaxation rate should monotonously increase with decreasing 
$E_{\rm Z}$ and remain small even at $E_{\rm Z}=0$.

%%%%%%%%%%%%%%%%%%%%%%%%%%%%%%%%%%%%%%%%%%%%%%%%%%%%%%%%%%%%%%%%%%%%%
\section{Fractional quantum Hall regime}
%%%%%%%%%%%%%%%%%%%%%%%%%%%%%%%%%%%%%%%%%%%%%%%%%%%%%%%%%%%%%%%%%%%%%

Recent NMR\cite{Kuzma98} and optical\cite{Davies98} experiments near 
$\nu={1\over3}$ revealed similar dependence of the electron spin 
polarization on the magnetic field to that found earlier at $\nu=1$.
In contrast to an earlier prediction,\cite{Kamilla96} it now seems 
plausible that the fast and weakly temperature-dependent nuclear 
spin relaxation near $\nu={1\over3}$ is somehow related to the 
presence of skyrmions in the Laughlin liquid.

Near $\nu=(2p+1)^{-1}$ ($p\ge1$ is an integer) Laughlin correlations
\cite{Laughlin83} allow mapping\cite{Jain89} of the low-energy 
interacting electron states onto the noninteracting composite 
fermion (CF) states with an effective filling factor $\nu^*\approx1$.
The Chern-Simons transformation, in which $2p$ magnetic flux quanta
are attached to each electron, results in the effective CF LL
degeneracy of $g^*=g-2p(N-1)$.
On a sphere,\cite{Haldane83} this replaces the electronic single 
particle angular momentum $l=Q\approx{1\over2}(2p+1)(N-1)$ by an
effective CF angular momentum $l^*=Q^*\approx{1\over2}(N-1)$, 
where $2Q^*$ denotes the effective CF monopole strength.

There are two types of low energy charge-neutral excitations of
Laughlin $\nu={1\over3}$ ground state, similar to the charge and 
spin waves\cite{Kallin84} of the $\nu=1$ state.
This similarity lies at the heart of the CF picture,\cite{Jain89}
where these excitations correspond to promoting one CF from 
a completely filled lowest ($n=0$) spin-$\downarrow$ CF LL either 
to the first excited ($n=1$) CF LL of the same spin ($\downarrow$) 
or to the same CF LL ($n=0$) but with the reversed spin ($\uparrow$).
Similarly to $\nu=1$, charge and spin waves at $\nu={1\over3}$ are 
composed of three types of elementary quasiparticles: a hole in the 
$n=0$ spin-$\downarrow$ CF LL and the particles in the $n=1$ 
spin-$\downarrow$ and $n=0$ spin-$\uparrow$ CF LL's, representing 
the Laughlin quasihole (QH) and quasielectron (QE) and the 
reversed-spin quasielectron (QE$_{\rm R}$), respectively.
Each of these quasiparticles is characterized by such single-particle
quantities as (fractional) electric charge, energy, or Landau
degeneracy of the single-particle Hilbert space.

\begin{figure}[t]
\epsfxsize=3.40in
\epsffile{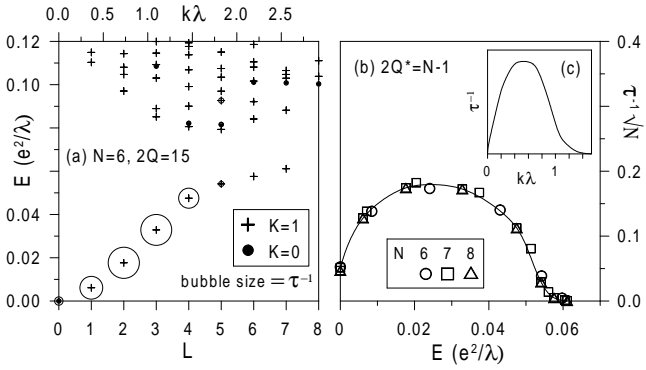}
\caption{
   Same as Fig.~\protect\ref{fig1} but for $\nu={1\over3}$.
   More excited states appear in frame (a) in addition to the 
   spin wave.}
\label{fig5}
\end{figure}

Expecting similar behavior, we have carried out similar calculations 
for the $\nu\approx{1\over3}$ filling as described in the previous 
section for $\nu\approx1$.
Let us begin with the Laughlin incompressible FQH state at precisely
$\nu={1\over3}$.
We found that the reversal of a nuclear spin in this state creates 
a spin wave, in perfect analogy to what happened in the IQH regime.
The main difference is that the spin wave at $\nu={1\over3}$ consists
of a QE$_{\rm R}$--QH pair whose interaction energy scale is about an 
order of magnitude smaller than it was for a $e_{\rm R}$--$h$ pair at 
$\nu=1$, predominantly due to the fractional QE$_{\rm R}$ and QH charge
(but also due to a larger size of the QE$_{\rm R}$ and QH wave functions).
In Fig.~\ref{fig5} we show the graphs for the FQH regime similar to 
those of Fig.~\ref{fig1}.
The spectrum shown in Fig.~\ref{fig5}(a) is for $N=6$ electrons at 
$2Q=15$.
In the CF picture, this corresponds to six CF's filling exactly their 
lowest spin-$\downarrow$ LL of degeneracy $g^*=2Q^*+1=6$.
Similarly to $\nu=1$, additional weak transitions to higher states 
appear in a small system, but the spin wave will remain the dominant 
feature of the spectrum in the $N\rightarrow\infty$ limit.

In Fig.~\ref{fig5}(b) we overlay the spin wave spectra $\tau^{-1}(E)$ 
obtained for different values of $N$.
Similarly to $\nu=1$, all data points (with $\tau^{-1}$ multiplied
by $\sqrt{N}$) fall on the same regular curve which, as expected,
vanishes in both $E=0$ and $E=\infty$ limits, and reaches maximum at
the energy $E\approx0.025\,E_{\rm C}$, about an order of magnitude 
smaller than at $\nu=1$.
In Fig.~\ref{fig5}(c) we replot $\tau^{-1}$ as a function of wave 
vector $k=L/R$.
By analogy to $\nu=1$, we expect that the length of spin wave most 
strongly coupled by $F$ to a nuclear spin reversal corresponds to 
the smallest area containing one electron (and thus containing one
unit of electron spin that must flip to compensate the nuclear spin).
For the uniform $\nu={1\over3}$ state, the average area per electron 
is 3 times larger than at $\nu=1$, yielding $\sqrt{3}$ times larger
length scale $\xi$, and thus the maximum of $\tau^{-1}(k)$ is expected 
at $k=\xi^{-1}\sim(\sqrt{3}\lambda)^{-1}\approx0.58\lambda^{-1}$.
Indeed, this seems to be true of our Fig.~\ref{fig5}(c).

To summarize our results at precisely $\nu={1\over3}$, the mechanism 
of the coupling to the nuclear spins is very analogous.
The two major differences can be predicted from the simple arguments
of 3 times reduced electron density and 3 times reduced charge of the 
involved quasiparticles.
These differences are (i) about $\sqrt{3}$ times larger characteristic 
length scale of the response $\xi$ (wave length of the ``active''
spin wave) and (ii) about $3^2$ smaller interaction energy $E_{\rm SW}
(\xi^{-1})$ of such active spin wave.
While (i) is the reason for the {\em reduction} of $\tau^{-1}$ 
of the corresponding transitions (compare the maxima of $\sqrt{N}
\tau^{-1}$ in Figs.~\ref{fig1} and \ref{fig5}), 
(ii) should actually {\em enhance} nuclear relaxation due to spin 
waves (if $E_{\rm Z}$ can be made disappear)
as a result of the weaker violation of the energy conservation.

Let us now turn to the spin-flip processes involving FQH skyrmions.
Their energy spectra and gaps for the ``internal'' excitations are
very similar to those at $\nu=1$ except for an overall reduction of 
the interaction energy scale and breaking of the skyrmion--antiskyrmion
symmetry.\cite{skyrmion}
In particular, graphs analogous to those in Fig.~\ref{fig2} describe
also the FQH skyrmions, only with about an order of magnitude smaller 
critical values of $E_{\rm Z}$.

\begin{figure}[t]
\epsfxsize=3.40in
\epsffile{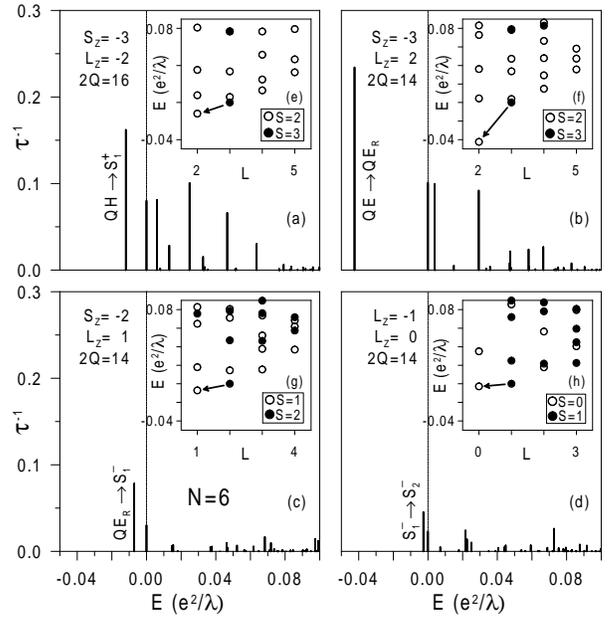}
\caption{
   Same as Fig.~\protect\ref{fig3} but for $\nu={1\over3}$.
   The initial states in different frames are:
   QH (a), QE (b), QE$_{\rm R}$ (c), and $S_1^-$ (d).}
\label{fig6}
\end{figure}

In Fig.~\ref{fig6} we display some of the $\tau^{-1}(E)$ spectra 
calculated for $N=6$ electrons at $2Q^*=N$ and $N+2$, corresponding 
to one skyrmion (or QE, or QE$_{\rm R}$) and antiskyrmion (or QH) 
in the Laughlin $\nu={1\over3}$ state, respectively.
As in Fig.~\ref{fig3}, in the insets we show the energy spectra
in which the initial and final states for each transition can
be found. 
In perfect analogy to the IQH system, we identify the sequences 
of $F$-induced transitions with $E<0$ that occur between the 
skyrmion or antiskyrmion states of different $K$: ${\rm QE}
\rightarrow{\rm QE}_{\rm R}\rightarrow S_1^-\rightarrow S_2^-
\rightarrow\dots$ and ${\rm QH}\rightarrow S_1^+\rightarrow 
S_2^+\rightarrow\dots$.

%%%%%%%%%%%%%%%%%%%%%%%%%%%%%%%%%%%%%%%%%%%%%%%%%%%%%%%%%%%%%%%%%%%%%
\section{Comparison of IQH and FQH regimes}
%%%%%%%%%%%%%%%%%%%%%%%%%%%%%%%%%%%%%%%%%%%%%%%%%%%%%%%%%%%%%%%%%%%%%

\begin{figure}[t]
\epsfxsize=3.40in
\epsffile{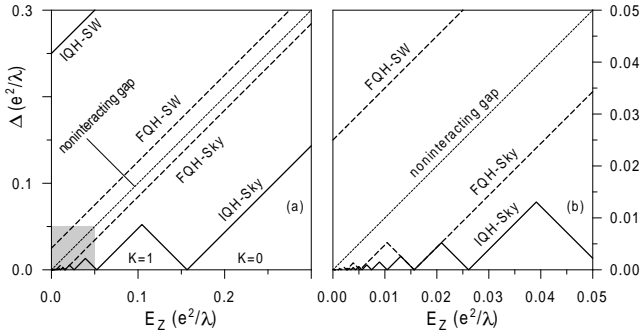}
\caption{
   The comparison of transition energies $\Delta$ of IQH and FQH 
   systems corresponding to the spin wave emission and the internal 
   skyrmion excitations, obtained from Eq.~(\ref{eqES}).
   $E_{\rm Z}$ is the Zeeman energy and $\lambda$ is the magnetic 
   length. 
   Frame (b) shows a blow up of the shaded part of frame (a).}
\label{fig7}
\end{figure}

Let us finally compare the skyrmion and spin wave excitations 
and their possible coupling to the nuclear spins in the IQH 
and FQH regimes.
It turns out that the following behavior depicted in Fig.~\ref{fig7} 
that can be predicted from the simple arguments and existing 
analytical results alone is not far from the results of our exact 
calculations.
The skyrmion energy spectrum is adequately reproduced by 
Eq.~(\ref{eqES}), with ${\cal E}={1\over4}\sqrt{\pi/2}\,E_{\rm C}$ 
for the IQH regime\cite{Sondhi93} and an about 10 times smaller 
value for the FQH regime.
This determines the dependence of the skyrmion size $K$ and 
its gap $\Delta$ for spin excitations ($S_K\rightarrow S_{K\pm1}$) 
on $E_{\rm Z}$ in both regimes.
These plots of $\Delta(E_{\rm Z})$ are marked as ``IQH-Sky'' 
and ``FQH-Sky'' in Fig.~\ref{fig7}.
On the other hand, it follows from the fact that $\lambda$ is the 
size of a cyclotron orbit that the spin waves must have $k\sim
\lambda^{-1}$ to strongly couple to a localized spin reversal
at $\nu=1$.
And knowing\cite{Kallin84} the spin wave dispersion $E_{\rm SW}
(k)$ allows an estimate of the total (Zeeman plus Coulomb) 
energy gap for such ``active'' IQH spin waves, $\Delta\sim 
E_{\rm Z}+{1\over4}E_{\rm C}$.
By reducing the interaction energy by an order of magnitude one 
can also predict the spin wave gap in the FQH regime, $\Delta
\sim E_{\rm Z}+{1\over40}E_{\rm C}$.
These two plots of $\Delta(E_{\rm Z})$ are marked as ``IQH-SW'' 
and ``FQH-SW'' in Fig.~\ref{fig7}.

Since the energy conservation requires that $\Delta$ be equal or 
at least close to the nuclear Zeeman energy which is essentially 
zero, it is clear from Fig.~\ref{fig7} how the relative efficiencies 
of spin waves and skyrmions in both regimes depend on $E_{\rm Z}$.
It is noteworthy that the ($e_{\rm R}$--$h$ or QE$_{\rm R}$--QH) 
interactions can have different effect on the spin gap $\Delta$, 
depending on the presence of skyrmions in the system.
Compared to a noninteracting system for which $\Delta=E_{\rm Z}$,
the interactions {\em increase} the spin gap $\Delta$ associated 
with the creation of spin waves, but {\em decrease} such a gap
associated with the skyrmion excitations.
For spin waves, the enhancement of $\Delta$ is due to a decrease 
of $e_{\rm R}$--$h$ or QE$_{\rm R}$--QH attraction at a finite 
wave vector $k\sim\sqrt{\nu}/\lambda$ (that can be interpreted
as a spin wave kinetic energy).
For skyrmions, the reduction of $\Delta$ is due to interaction
induced level crossings and ground state transitions.

It is clear from Fig.~\ref{fig7} that the strong interactions in 
the IQH regime prevent the spin waves at $\nu=1$ from coming into 
resonance with nuclear spins regardless of the value of $E_{\rm Z}$ 
and practically eliminate them as an efficient nuclear spin relaxation 
mechanism at this filling.
But at the same time, these interactions allow efficient relaxation
near $\nu=1$ through the spin-flip nucleus--skyrmion collisions
over a long range of $E_{\rm Z}$.
On the other hand, the much weaker interactions in the FQH regime
do not completely exclude nuclear relaxation by means of spin wave 
emission at $E_{\rm Z}\approx0$, but they considerably shorten the range 
of $E_{\rm Z}$ in which the skyrmions occur and can spin-flip collide 
with the nuclei.
Therefore, a drop of the nuclear spin relaxation time $\tau$ caused 
by the introduction of charge excitations to the incompressible liquid
(by varying density or magnetic field to move $\nu$ away from 1 or 
${1\over3}$, increasing temperature, or inducing current) should be 
more pronounced in the IQH regime.
This agrees with the experiments that typically show much longer 
relaxation times at $\nu=1$ than at $\nu={1\over3}$.

%%%%%%%%%%%%%%%%%%%%%%%%%%%%%%%%%%%%%%%%%%%%%%%%%%%%%%%%%%%%%%%%%%%%%
\section{Conclusion}
%%%%%%%%%%%%%%%%%%%%%%%%%%%%%%%%%%%%%%%%%%%%%%%%%%%%%%%%%%%%%%%%%%%%%

Using exact numerical diagonalization techniques we have studied
possible relaxation mechanisms of nuclear spins coupled through 
the hyperfine interaction to the quantum Hall states of a 2DEG at 
filling factors near $\nu=1$ and ${1\over3}$.
By extrapolation of our finite-size results, we were able to determine 
the spectral function $\tau^{-1}(E)$ describing response of an infinite 
(planar) 2DEG to the reversal of an embedded localized spin.
We found that the spectral function can be decomposed into a continuous 
part describing transitions from the incompressible ``background'' 
state, and a discrete part which is due to the presence of additional
charge excitations (skyrmions).

The continuous part of the response function $\tau^{-1}(E)$ describes 
emission of a spin wave, whose energy $E_{\rm SW}$ is a sum of the 
electronic Zeeman gap $E_{\rm Z}$ and the kinetic energy dependent 
on the wave vector $k$.
We found that, when expressed as a function of wave vector, 
$\tau^{-1}$ vanishes in both $k=0$ and $\infty$ limits, and that 
it has a maximum at a finite $k\approx\lambda^{-1}\sqrt{\nu}$.
Using the known spin wave dispersion for $\nu=1$, we showed that 
the emission energy of an ``active'' spin wave that can couple to 
a nuclear spin reversal exceeds $E_{\rm Z}$ by a kinetic term 
$\sim{1\over4}E_{\rm C}\cdot\nu^2$.
This implies that even in the limit of vanishing $E_{\rm Z}$, 
the energy conservation law will prevent coupling of the (locally) 
incompressible quantum Hall states to nuclear spins.
This result agrees with long nuclear spin relaxation times observed 
in experiments in the absence of charged excitations (at precisely 
$\nu=1$ or ${1\over3}$ and at low temperature).

The situation changes dramatically when skyrmions are introduced 
into the 2DEG by either moving $\nu$ away from 1 or ${1\over3}$, 
increasing temperature, or applying voltage to induce electric current.
The reason for a different behavior are the so-called ``internal'' 
spin excitations of skyrmions (in which their spin $K$ changes by one) 
whose energy is much smaller than $E_{\rm Z}$ and can be brought into 
resonance with the nearly vanishing nuclear Zeeman energy.
Moreover, we have checked that the oscillator strength $\tau^{-1}$ 
of the skyrmion--nucleus collision corresponding to the $K\leftrightarrow
(K+1)$ transition is large and increases with increasing $K$.

In both IQH and FQH regimes, our results imply critical dependence 
of the nuclear spin relaxation rate on the presence of skyrmions 
in the 2DEG, in good agreement with the experiments.
The contrast between the forbidden relaxation due to spin waves
and the efficient relaxation due to skyrmions should be more 
pronounced at $\nu=1$ than at $\nu={1\over3}$ because of a larger 
(by an order of magnitude) interaction energy scale.

%%%%%%%%%%%%%%%%%%%%%%%%%%%%%%%%%%%%%%%%%%%%%%%%%%%%%%%%%%%%%%%%%%%%%
\section{Acknowledgment}
%%%%%%%%%%%%%%%%%%%%%%%%%%%%%%%%%%%%%%%%%%%%%%%%%%%%%%%%%%%%%%%%%%%%%

The authors acknowledge partial support by the Materials Research
Program of Basic Energy Sciences, US Department of Energy.

%%%%%%%%%%%%%%%%%%%%%%%%%%%%%%%%%%%%%%%%%%%%%%%%%%%%%%%%%%%%%%%%%%%%%

\end{document}